\documentclass[final]{aipproc}
\layoutstyle{6x9}

\usepackage{amsmath,bm}
\usepackage{graphics}

\allowdisplaybreaks[1]
\def\be#1\ee{\begin{equation}#1\end{equation}}
\def\bal#1\eal{\begin{align}#1\end{align}}
\newcommand{\n}{\notag}

\newcommand{\abs}[1]{\lvert#1\rvert}
\newcommand{\supfi}[1]{{}^{\,#1\!}}

\renewcommand{\bf}{\mathbf}
\renewcommand{\cal}{\mathcal}

\begin{document}

\title{Muonium spectrum beyond the \\
nonrelativistic limit}


\author{Axel Weber}{
address={Instituto de F\'{\i}sica y Matem\'aticas, Universidad Michoacana de
San Nicol\'as de Hidalgo,\\
Edificio C-3, Ciudad Universitaria, A.P. 2-82, 58040 Morelia, Michoac\'an , 
Mexico}}

\begin{abstract}
A generalization of the Gell-Mann--Low theorem is applied to the
antimuon-electron system. The bound state spectrum is extracted numerically.
As a result, fine and hyperfine structure are reproduced correctly near
the nonrelativistic limit (and for arbitrary masses). We compare the spectrum
for the relativistic value $\alpha = 0.3$ with corresponding calculations 
in light-front quantization.
\end{abstract}

\maketitle

More than thirty years have passed since quantum chromodynamics (QCD) has
been formulated. Overwhelming evidence has accumulated during this time that 
QCD gives a complete description of hadronic physics at every energy scale 
currently
accessible to experiment. However, the status of the theory is not at all
satisfactory: we are still searching for a deeper understanding of the 
interaction between quarks and gluons that arises from the QCD Lagrangian
beyond the high momentum transfer limit where perturbation theory is
applicable. Parts of the physical spectrum that is generated by the theory
can be determined through lengthy numerical calculations on space-time
lattices which, however, do not provide major insights into the
underlying dynamics. Ideally, one would like to have an analytical
description of the interaction between quarks and gluons derived from the
QCD Lagrangian, typically in the form of an approximation and a
systematic procedure to incorporate corrections to the latter. No such 
description is available to date. Moreover, once it is obtained,
we still face the problem of calculating the bound states resulting from
this interaction, i.e., the physical hadrons.

The present contribution deals with this second step in connecting the
theory in a transparent way with the phenomenology. Without an
appropriate description of the fundamental interaction at hand, we 
consider a simpler theory, namely, quantum electrodynamics (QED), which 
still has some features in common with QCD. In order to simulate a
relativistic situation like the one that prevails in the lighter hadrons,
we artificially consider larger coupling constants. E.g., $\alpha
= 0.3$ was used before in  light-front calculations we will compare with 
later on. For concreteness, we will consider muonium, a bound state
of an antimuon with an electron, although we will allow for an arbitrary 
antimuon mass. In the special case of equal masses of electron and
``antimuon'', the system is similar to positronium if we disregard all
(virtual and real) annihilation processes there.

Our approach to relativistic bound states is an application of
a generalization of the Gell-Mann--Low theorem \cite{Web00} to the
subspace of Fock space that contains all states of one electron and one
antimuon (and no photon). As in the earlier applications of the same formalism
to the Wick-Cutkosky model and Yukawa theory \cite{WL02, WL05}, the effective 
Hamiltonian generated by the generalized Gell-Mann-Low theorem contains the 
relativistic kinetic energies of the constituents
and an effective potential. To lowest order in a perturbative
expansion, the matrix elements of the effective potential read, in Coulomb
gauge,
\bal
\lefteqn{\langle \bf{p}_A, r; \bf{p}_B, s | V_{\text{eff}} 
| \bf{p}'_A, r'; \bf{p}'_B, s' \rangle =
- \frac{e^2}{\sqrt{2 E_{\bf{p}_A}^A \, 2 E_{\bf{p}_B}^B
\, 2 E_{\bf{p}_A'}^A \, 2 E_{\bf{p}_B'}^B}}} \hspace{2cm} \n \\
&\times \Bigg[ \frac{1}{\left( \bf{p}_A - \bf{p}_A' \right)^2}
\left[ \bar{u}_A (\bf{p}_A, r) \gamma^0 u_A (\bf{p}_A', r') \right] 
\left[ \bar{u}_B (\bf{p}_B, s) \gamma^0 u_B (\bf{p}_B', s') \right] \n \\
&\phantom{\times} - \frac{1}{2 \abs{\bf{p}_A - \bf{p}_A'}} \left( \frac{1}{
E_{\bf{p}_A}^A + \abs{\bf{p}_A - \bf{p}_A'} - E_{\bf{p}_A'}^A} + \frac{1}{
E_{\bf{p}_B}^B + \abs{\bf{p}_B - \bf{p}_B'} - E_{\bf{p}_B'}^B} \right) \n \\
&\phantom{\times -} \times \left[ \bar{u}_A (\bf{p}_A, r) 
\gamma^i u_A ({\mathbf p}_A', r') \right] 
\left( \sum_{\lambda = 1}^2 \varepsilon_i^{(\lambda)} 
(\bf{p}_A - \bf{p}_A') \, \varepsilon_j^{(\lambda) \ast} 
(\bf{p}_A - \bf{p}_A') \right) \n \\
&\phantom{\times -} \hspace{1.5cm} \times \left[ \bar{u}_B (\bf{p}_B, s) 
\gamma^j u_B (\bf{p}_B', s') \right] \Bigg] (2 \pi)^3 \delta 
(\bf{p}_A + \bf{p}_B - \bf{p}'_A - \bf{p}'_B) \:. \label{HBQED}
\eal
Here, $| \bf{p}_A, r; \bf{p}_B, s \rangle$ symbolizes the state of an 
electron with 3-momentum $\bf{p}_A$ and spin orientation $r$ (in a spinor 
basis yet to be specified) and an antimuon with 3-momentum $\bf{p}_B$
and spin orientation $s$. We use the shorthands
$E_{\bf{p}_A}^A = (m_A^2 + \bf{p}_A^2)^{1/2}$ and 
$E_{\bf{p}_B}^B = (m_B^2 + \bf{p}_B^2)^{1/2}$
for the kinetic energies. For convenience, we have introduced the
charge-conjugate Dirac spinors $u_B (\bf{p}_B, s)$ for the antimuon,
while $u_A (\bf{p}_A, s)$ represents the electron spinors.
The spatially transverse photon polarization vectors 
$\varepsilon^{(\lambda)}_i (\bf{k})$ satisfy the relation
$\sum_{\lambda = 1}^2 \varepsilon_i^{(\lambda)} (\bf{k}) \, 
\varepsilon_j^{(\lambda) \ast} (\bf{k}) = \delta^{\text{tr}}_{ij} (\bf{k}) 
= \delta_{i j} - \hat{k}_i \hat{k}_j$ (where 
$\hat{\bf{k}} = \bf{k}/\abs{\bf{k}}$).

As for the interpretation of the effective potential \eqref{HBQED}, the
second line stems from the instantaneous Coulomb potential, easily
identified by the momentum dependence in the denominator (the Fourier
transform of the spatial Coulomb potential), and multiplied with the charge
densities of the Dirac currents. The following lines are the result of
transverse photon exchange, the more complicated denominator indicating
a retarded interaction, and the Dirac currents being contracted with the
corresponding photon polarization vectors. 

The delta function in Eq.\ \eqref{HBQED} shows that total 3-momentum
is conserved by the effective interaction, and in the following we will 
consider the center-of-mass system (c.m.s.) $\bf{p}_A + \bf{p}_B = \bf{p}'_A 
+ \bf{p}'_B = 0$. In order to simplify the diagonalization of the
effective Hamiltonian, we express the
Dirac spinors in terms of Pauli spinors (using the Dirac-Pauli representation)
to find for the effective Schr\"odinger equation in the c.m.s.,
\bal
\lefteqn{\left( \sqrt{m_A^2 + \bf{p}^2} + \sqrt{m_B^2 + \bf{p}^2} \right)
\phi (\bf{p}) - e^2 \int \frac{d^3 p'}{(2 \pi)^3} \sqrt{
\frac{E^A_{\bf{p}} + m_A}{2 E^A_{\bf{p}}} \,
\frac{E^B_{\bf{p}} + m_B}{2 E^B_{\bf{p}}} \,
\frac{E^A_{\bf{p}'} + m_A}{2 E^A_{\bf{p}'}} \,
\frac{E^B_{\bf{p}'} + m_B}{2 E^B_{\bf{p}'}}}} \n \\
&\times \Bigg[ \frac{1}{\left( \bf{p} - \bf{p}' \right)^2}
\left( 1 + \frac{\bf{p} \cdot \bm{\sigma}_A}{E^A_{\bf{p}} + m_A} \,
\frac{\bf{p}' \cdot \bm{\sigma}_A}{E^A_{\bf{p}'} + m_A} \right)
\left( 1 + \frac{\bf{p} \cdot \bm{\sigma}_B}{E^B_{\bf{p}} + m_B} \,
\frac{\bf{p}' \cdot \bm{\sigma}_B}{E^B_{\bf{p}'} + m_B} \right) \n \\
& + \frac{1}{2 \abs{\bf{p} - \bf{p}'}} \left( \frac{1}{E_{\bf{p}}^A +
\abs{\bf{p} - \bf{p}'} - E_{\bf{p}'}^A} + \frac{1}{E_{\bf{p}}^B +
\abs{\bf{p} - \bf{p}'} - E_{\bf{p}'}^B} \right) \left( 
\frac{(\bf{p} \cdot \bm{\sigma}_A) \sigma_A^i}{E^A_{\bf{p}} + m_A}
+ \frac{\sigma_A^i (\bf{p}' \cdot \bm{\sigma}_A)}{E^A_{\bf{p}'} + m_A} 
\right)\n \\
&\phantom{\times +} \hspace{2.5cm} 
\times \delta^{\text{tr}}_{i j} (\bf{p} - \bf{p}')
\left( \frac{(\bf{p} \cdot \bm{\sigma}_B) \sigma_B^j}{E^B_{\bf{p}} + m_B}
+ \frac{\sigma_B^j (\bf{p}' \cdot \bm{\sigma}_B)}{E^B_{\bf{p}'} + m_B} \right) 
\Bigg] \phi (\bf{p}') = E' \phi (\bf{p}) \:. \label{schrQED}
\eal
Here, $\bf{p} = \bf{p}_A = - \bf{p}_B$, the spinorial wave function 
$\phi (\bf{p})$ is defined as
$\phi (\bf{p}_A) (2 \pi)^3 \delta (\bf{p}_A + \bf{p}_B) 
= \sum_{r,s} \langle \bf{p}_A, r; \bf{p}_B, s | \phi \rangle 
\left[ \chi_r \otimes \chi_s \right]$,
and $\bm{\sigma}_A$ ($\bm{\sigma}_B$) is understood to act on the Pauli
spinor $\chi_r$ ($\chi_s$) only. $E'$ is the difference between the
energy of the bound state and the vacuum energy.
Of the full state $| \phi \rangle$ in Fock space (with zero total momentum), 
only its projection to one-electron--one-antimuon states
$| \bf{p}_A, r; \bf{p}_B, s \rangle$ appears. The effect of its components
in other Fock space sectors is taken care of implicitly by the effective
potential.

In order to solve the Schr\"odinger equation \eqref{schrQED}, we take into 
account its rotational symmetry. Eigenstates of total angular momentum $J$ 
can be constructed as usual by adding relative orbital angular momentum 
$L$ and total spin $S$. For convenience, we will label the eigenstates by 
the ``relative parity'' $\pi'$ defined through $(-1)^L = \pi' (-1)^J$.
Since $S=0,1$, for given $J$ the sector $\pi'=+1$ contains the states
with $L=J$ and $S=0$ or $S=1$, while for $\pi'=-1$ we can have $L=J-1$ or
$L=J+1$, with $S=1$ in both cases. In any sector $J^{\pi'}$, the two different
possible $(LS)$-states will mix, except in the following cases: (i) for $J=0$, 
$J^{\pi'} = 0^+$ is only realized by $(L=0, S=0)$, and $0^-$ only by 
$(L=1, S=1)$; (ii) in the case of equal masses, the Hamiltonian acquires an
additional symmetry under the exchange of particles $A$ and $B$; as a
result, $S$ becomes a good quantum number and there is no mixing in the
$(\pi'=+1)$-sector; (iii) in the one-body limit where one of the masses goes
to infinity, the spin of the heavy particle decouples from the dynamics;
as a result, every two states are degenerate in this limit and $L$ becomes 
a good quantum number [no mixing in the $(\pi'=-1)$-sector].

After explicitly carrying out the contractions of the spatial indices in 
the transverse photon exchange part, the formulae derived before for the 
application to Yukawa theory \cite{WL05} can be used to determine the
result of the application of the terms containing the Pauli matrices in
Eq.\ \eqref{schrQED} to the total angular momentum eigenstates. On the other 
hand, the application of the factors containing $\abs{\bf{p} - \bf{p}'}$
on orbital angular momentum eigenstates proceeds through the partial wave
decomposition of the former. The partial wave decomposition of the
(Fourier transformed) Coulomb potential is well-known, the one for the
$\delta_{ij}$-part of the transverse photon exchange has been calculated
in Ref.\ \cite{WL05}. The partial waves of the $\hat{k}_i \hat{k}_j$-part
of the transverse photon exchange are given by
\bal
b_L (p, p') &= \frac{2L + 1}{2} \int_{-1}^1 d \cos \theta \, P_L (\cos \theta)
\n \\[-2mm]
&\phantom{=} \times \frac{1}{\left( \bf{p} - \bf{p}' \right)^2} 
\, \frac{1}{2 \abs{\bf{p} - \bf{p}'}}
\left( \frac{1}{E_p^A + \abs{\bf{p} - \bf{p}'} - E_{p'}^A} + 
\frac{1}{E_p^B + \abs{\bf{p} - \bf{p}'} - E_{p'}^B} \right) \:.
\label{partwaveb}
\eal
This integral diverges like $(p - p')^{-2}$ for $p' \to p$ which would lead
to a divergence in the $p'$-integral. These divergences, of course, are
spurious and cancel in pairs. However, for the numerical calculation,
we have to extract the divergent parts and perform the cancellations
analytically. Fortunately, the extraction of the divergencies is simple: they 
occur at $\cos \theta = 1$, and since $P_L (1) = 1$, we can define
``reduced'' Legendre polynomials $P_L^R (\cos \theta)$ through
\be
P_L (\cos \theta) - 1 =  \left( \cos \theta - 1 \right) P_L^R (\cos \theta) 
\:. \label{defPR}
\ee
Separating the one on the l.h.s.\ of Eq.\ \eqref{defPR}
under the integrals \eqref{partwaveb}, the remainder of the integrals is
logarithmically divergent for $p' \to p$ (as are the other partial waves), 
and the following $p'$-integration is convergent. The divergent
parts in Eq.\ \eqref{partwaveb} originating from the one in Eq.\ \eqref{defPR}
can be analytically cancelled in pairs, leaving a finite contribution.

Putting everything together, the potential term in the Schr\"odinger 
equation \eqref{schrQED} reduces to a one-dimensional integral over $p'$
when applied to the total angular momentum eigenstates. For every sector
$J^{\pi'}$, two such equations are coupled (with the exceptions mentioned
above). The explicit expressions for the integral kernels (diagonal and 
off-diagonal because of the coupling) are quite lengthy and cannot be
reproduced here due to lack of space.

\begin{figure}
\resizebox{\textwidth}{!}
  {\rotatebox{-90}{\includegraphics{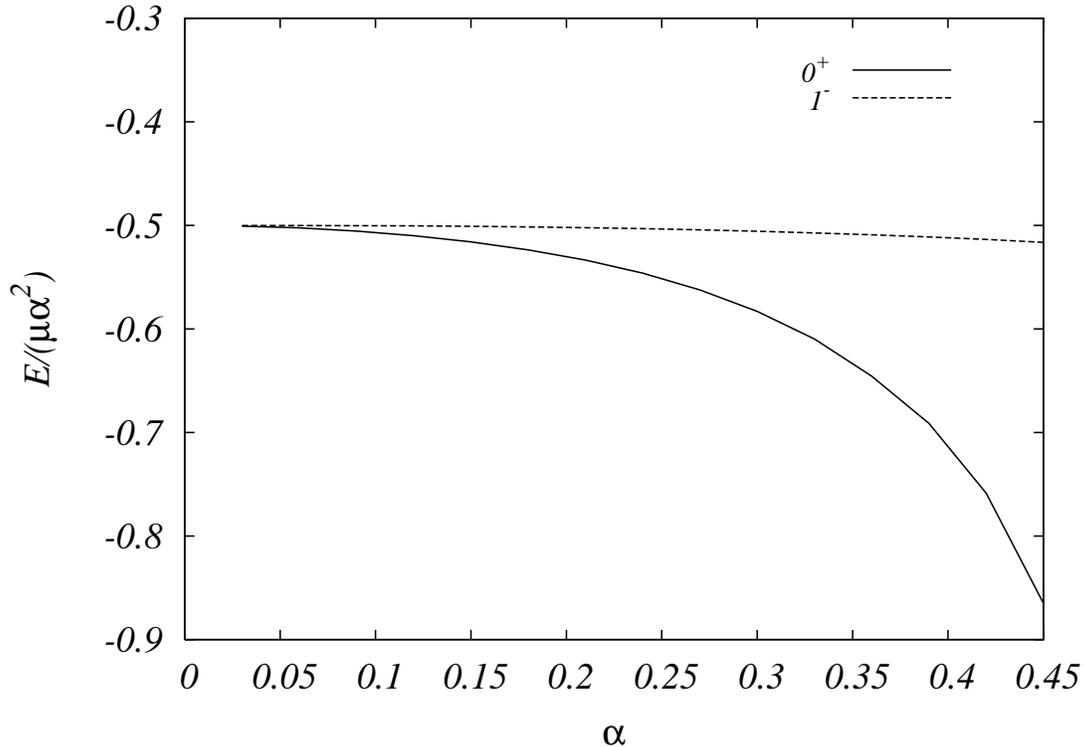}}}
\caption{The binding energy $E = E' - m_A - m_B$ as a function of the
fine structure constant $\alpha = e^2/4 \pi$ for the case of equal masses
$m_A = m_B$. $E$ is normalized to $\mu \alpha^2$ where $\mu$ is the
reduced mass ($\mu = m_A/2$ in the present case). Plotted are the lowest
energy levels for $J^{\pi'} = 0^+$ and $1^-$ corresponding to the
nonrelativistic principal quantum number $n=1$. \label{fig1QED}}
\end{figure}

\begin{figure}
\resizebox{\textwidth}{!}
  {\rotatebox{-90}{\includegraphics{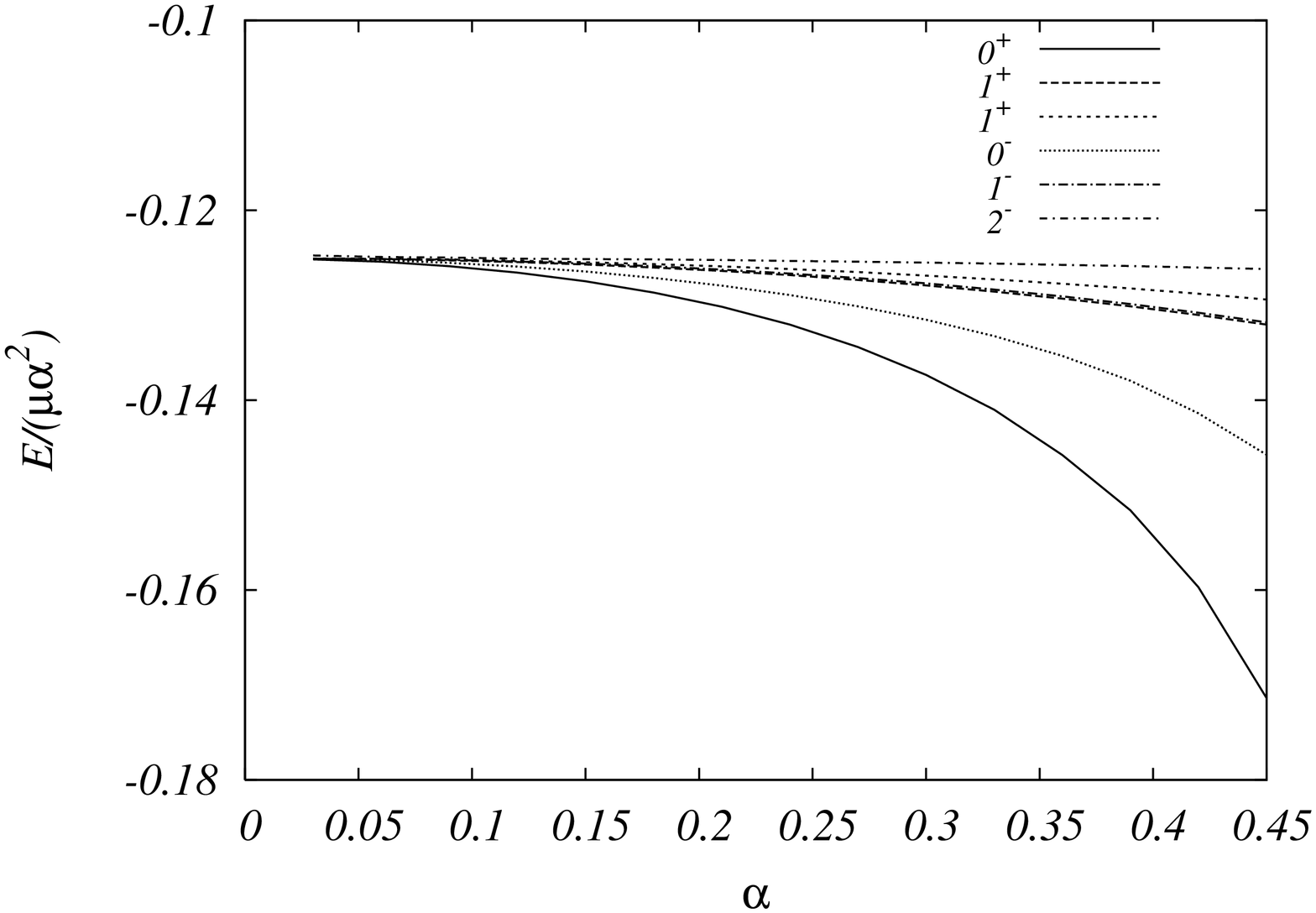}}}
\caption{As Fig.\ \ref{fig1QED}, but for the energy levels corresponding
to the nonrelativistic principal quantum number $n=2$. \label{fig2QED}}
\end{figure}

The (coupled) one-dimensional integral equations can be solved by
expanding the wave function in an appropiate orthonormal basis. After 
reducing the basis to a finite number of elements (40 in our calculations), 
the integral equations are approximately replaced by finite matrix equations.
The matrix elements are two-dimensional integrals which are calculated
numerically (we use a two-dimensional grid of $400 \times 800$ points). 
Finally, the matrices are numerically diagonalized to give the (approximate)
eigenvalues and eigenstates of Eq.\ \eqref{schrQED}.
The results for the lowest energy eigenvalues are plotted in Figs.\
\ref{fig1QED} and \ref{fig2QED} for equal constituent masses and fine 
structure constants $\alpha \leq 0.45$. 
The binding energies are normalized to $\mu \alpha^2$, $\mu$ being the
reduced mass, so that the comparison with the nonrelativistic energy
eigenvalues $\mu \alpha^2/2 n^2$ is immediate.

We find that for values $\alpha < 0.1$, the energy levels are
dominated by the nonrelativistic values plus the
leading relativistic corrections (the leading-order
fine and hyperfine structure) of order $\mu \alpha^4$, both in our
numerical results and in the perturbative calculations of bound-state
QED. In this region of small coupling constants, the numerical results
are in good agreement with the perturbative calculations, apparently
only limited by the numerical precision. For larger values $\alpha
> 0.1$, higher perturbative orders become important and our
numerical results deviate in some cases strongly from the lowest-order
perturbative predictions. 

In Table \ref{tabQED} we compare our results for $\alpha = 0.3$ with two 
different calculations in light front quantization \cite{TP97, GPW98} (we 
use the data for the Gaussian similarity function in the latter paper).
\begin{table}
\begin{tabular}{ccccc}
\hline \tablehead{1}{c}{b}{state}
  & \tablehead{1}{c}{b}{perturbation \\ theory}
  & \tablehead{1}{c}{b}{our \\ results}
  & \tablehead{1}{c}{b}{Ref.\ \cite{TP97}} 
  & \tablehead{1}{c}{b}{Ref.\ \cite{GPW98}} \\
\hline
$1 \supfi{1}S_0 \, (0^+)$ & $-0.559$ & $-0.583$ & $-0.525$ & $-0.551$ \\
$1 \supfi{3}S_1 \, (1^-)$ & $-0.499$ & $-0.506$ & $-0.501$ & $-0.525$ \\
$2 \supfi{1}S_0 \, (0^+)$ & $-0.1343$ & $-0.1373$ & $-0.1301$ & $-0.1332$ \\
$2 \supfi{3}P_0 \, (0^-)$ & $-0.1306$ & $-0.1315$ & $-0.1335$ & $-0.1369$ \\
$2 \supfi{3}P_1 \, (1^+)$ & $-0.1278$ & $-0.1279$ & $-0.1298$ & $-0.1327$ \\
$2 \supfi{3}S_1 \, (1^-)$ & $-0.1268$ & $-0.1277$ & $-0.1269$ & $-0.1298$ \\
$2 \supfi{1}P_1 \, (1^+)$ & $-0.1268$ & $-0.1269$ & $-0.1290$ & $-0.1315$ \\
$2 \supfi{3}P_2 \, (2^-)$ & $-0.1255$ & $-0.1255$ & $-0.1277$ & $-0.1302$ \\
\hline
\end{tabular}
\caption{Binding energies $E/\mu \alpha^2$ for equal masses from 
perturbation theory to $\cal{O} (\mu \alpha^4)$, from our
numerical results, and from Refs.\ \cite{TP97} and \cite{GPW98}}
\label{tabQED}
\end{table}
In the table, we label the states by the nonrelativistic notation 
$n \supfi{2S+1}L_J$ and also indicate the corresponding sectors
$J^{\pi'}$. There is a clear tendency in our results towards more 
negative energies, i.e., stronger
binding, compared to $\cal{O} (\mu \alpha^4)$-perturbation theory. The
ordering of the different levels, however, is the same as in
perturbation theory. We can also see that the difference to perturbation 
theory in the direction of stronger binding is systematically larger for
$S$-states than for $P$-states, and also larger for $(J=0)$-states than for
$(J=1)$-states, and smallest for the $(J=2)$-state. For the light-front
results, this latter tendency is inverted; the $S_0$-states have
even higher energies than in perturbation theory. Both light-front
calculations are qualitatively similar, only that the binding
is stronger throughout in the similarity transform approach of Ref.\
\cite{GPW98}. In conclusion, we find very different results with the two 
different methods for relativistic bound state calculations (in
the approximations presently considered). We should remark, however, that
there is an unphysical logarithmic UV cutoff dependence in the light-front
results (for the cited values, the cutoff has been set equal to the
constituent masses).

\begin{theacknowledgments}
Support by Conacyt grant 46513-F and CIC-UMSNH is gratefully acknowledged. 
I thank my collaborator Juan Carlos L\'opez Vieyra for performing the 
numerical calculations and elaborating the graphics.
\end{theacknowledgments}


\begin{thebibliography}{9}

\bibitem{Web00} A. Weber, in \textit{Particles and Fields --- Seventh Mexican 
Workshop}, edited by A. Ayala, G. Contreras, and G. Herrera, AIP Conf.
Proc. No. 531, AIP, New York, 2000, pp. 305--309, preprint hep-th/9911198.

\bibitem{WL02} A. Weber, and N. E. Ligterink, \textit{Phys. Rev. D} 
\textbf{65}, 025009 (2002).

\bibitem{WL05} A. Weber, and N. E. Ligterink, ``Bound states in Yukawa 
theory'', preprint hep-ph/0506123.

\bibitem{TP97} U. Trittmann, and H.-C. Pauli, ``Quantum electrodynamics
at strong couplings'', preprint hep-th/9704215.

\bibitem{GPW98} E. L. Gubankova, H.-C. Pauli, F. J. Wegner, and G. Papp,
``Light-cone Hamiltonian flow for positronium'', preprint hep-th/9809143.

\end{thebibliography}
\end{document}